\documentclass[a4paper,11pt]
{article}

\usepackage{jheppub}

\usepackage{slashed}
\usepackage{wrapfig}
\usepackage{bm}
\usepackage{latexsym,amssymb,amsmath,float,url,mathrsfs}
\usepackage{latexsym}
\usepackage{graphicx}
\usepackage{epstopdf}
\usepackage{amsfonts}
\usepackage{amsmath}
\usepackage{amssymb}
\usepackage{comment}
\usepackage{subfigure}
\usepackage{natbib}
\usepackage{hyperref}
\usepackage{pifont}
\usepackage{blindtext}
\usepackage{adjustbox}
\usepackage{multirow}
\usepackage{tabularx}
\usepackage[table]{xcolor}
\usepackage{orcidlink}
\usepackage{tikz}
\usetikzlibrary{tikzmark}
\usepackage{amssymb}
\usepackage{bbold}
\usepackage{orcidlink}

\usepackage{blkarray}
\usepackage{graphicx}
\usepackage{amsfonts}
\usepackage{soul}
\usepackage{amssymb}
\usepackage{amsmath}
\usepackage{cancel}
\usepackage{tcolorbox}

\def\MSbar{\relax\ifmmode\overline{\rm MS}\else{$\overline{\rm MS}${ }}\fi}

\def\BH{{B\!-\!H}}
\def\bh{{\rm bh}}

\usepackage{graphics,appendix,afterpage,makecell} 

\definecolor{oucrimsonred}{rgb}{0.6, 0.0, 0.0}
\definecolor{persianblue}{rgb}{0.11, 0.22, 0.73}
\definecolor{forestgreen}{rgb}{0.13,0.35,0.13}
\definecolor{lightgray}{rgb}{0.83, 0.83, 0.83}
 \hypersetup{colorlinks, citecolor=oucrimsonred, linkcolor=black, urlcolor=oucrimsonred}
\definecolor{cornellred}{rgb}{0.7, 0.11, 0.11}
\definecolor{navyblue}{rgb}{0.0, 0.0, 0.5}
\definecolor{amethyst}{rgb}{0.6, 0.4, 0.8}
\definecolor{yellow}{rgb}{1.0, 1.0, 0.0}
\definecolor{firebrick}{rgb}{0.7, 0.13, 0.13}
\definecolor{tangerineyellow}{rgb}{1.0, 0.8, 0.0}
\definecolor{deepfuchsia}{rgb}{0.76, 0.33, 0.76}
\definecolor{amber}{rgb}{1.0, 0.75, 0.0}
\definecolor{VioletRed4}{rgb}{0.55, 0.13, .32}
\definecolor{indiagreen}{rgb}{0.07, 0.53, 0.03}
\definecolor{VioletRed4}{rgb}{0.55, 0.13, .32}
\newcommand{\be}{\begin{equation}}
\newcommand{\ee}{\end{equation}}
\newcommand{\bea}{\begin{equation} \begin{aligned}}
\newcommand{\eea}{\end{aligned} \end{equation}}

\definecolor{oucrimsonred}{rgb}{0.6, 0.0, 0.0}
\newcommand\vertarrowbox[3][6ex]{%
  \begin{array}[t]{@{}c@{}} #2 \\
  \left\uparrow\vcenter{\hrule height #1}\right.\kern-\nulldelimiterspace\\
  \makebox[0pt]{\scriptsize#3}
  \end{array}%
}

\definecolor{verdechiaro}{rgb}{0.6,1,0.6}
\definecolor{giallochiaro}{rgb}{1,1,0.6}
\definecolor{bluscuro}{rgb}{0.15, 0.2, 0.9}
\definecolor{verdes}{rgb}{0.1, 0.5, 0.1}%
\definecolor{tangerineyellow}{rgb}{1.0, 0.8, 0.0}

\definecolor{americanrose}{rgb}{1.0, 0.01, 0.24}
\definecolor{cobalt}{rgb}{0.0, 0.28, 0.67}
\definecolor{brandeisblue}{rgb}{0.0, 0.44, 1.0}
\definecolor{mycolor}{rgb}{0.0, 0.0, 0.5}
\definecolor{oxfordblue}{rgb}{0.0, 0.13, 0.28}
\definecolor{azure}{rgb}{0.0, 0.5, 1.0}
\definecolor{turquoiseblue}{rgb}{0.0, 1.0, 0.94}
\newtcolorbox{mynewbox}[1]{colback=white!5!white,colframe=azure!75!black,fonttitle=\bfseries,title=#1}
\newtcolorbox{mybox}{colback=mycolor!5!white,colframe=azure!75!black}
\newtcolorbox{mynamedbox}[1]{colback=mycolor!5!white,colframe=azure!75!black,title=#1}
\definecolor{venetianred}{rgb}{0.78, 0.03, 0.08}
\newtcolorbox{mynamedbox1}[1]{colback=venetianred!5!white,colframe=venetianred!80!black,title=#1}
\newtcolorbox{mynamedbox2}[1]{colback=azure!5!white,colframe=azure!80!black,title=#1}

\definecolor{verdes}{rgb}{0.1, 0.5, 0.1}%
\definecolor{cornellred}{rgb}{0.7, 0.11, 0.11}

\definecolor{VioletRed4}{rgb}{0.55, 0.13, .32}

\hypersetup{
     colorlinks   = true,
     citecolor    = violet,
     urlcolor     = violet,
     linkcolor    = violet}

\definecolor{rossocorsa}{rgb}{0.83, 0.0, 0.0}

\usepackage[normalem]{ulem}

\vspace*{1.5cm}

\title{ 
Information-Theoretic  Black Hole Entropy I: Beyond the Area Law}

\author[a]{Alex Kehagias\orcidlink{0000-0001-6080-6215}}
\affiliation[a]{Physics Division, National Technical University of Athens, Athens, 15780, Greece}

\abstract{Although the Bekenstein-Hawking area law is consistent with the first and second laws of black hole thermodynamics, it appears to be in conflict with the  third law, which in the Nernst formulation  states that the entropy should either vanish or approach a universal constant in the zero-temperature limit. We argue that this tension reflects a limitation of the semiclassical area law rather than a fundamental feature of black hole thermodynamics. On this basis, we obtain an entropy formula that is consistent with the third law, and approaches Bekenstein-Hawking entropy in the high-temperature limit.
The resulting entropy admits a simple microscopic interpretation, and it can be written as the Kullback-Leibler divergence between a mass-biased Bernoulli distribution and the uniform distribution on 
$N$ microscopic bits. From this perspective, the thermodynamic black hole entropy admits an
information-theoretic representation as an entropy deficit, namely as the
relative entropy between the black hole ensemble and a maximally mixed
reference ensemble. The Bekenstein-Hawking area law then emerges as the leading term in an $1/N$ expansion, while a universal mass scale $M_0=\sqrt{N}\,M_P$, interpreted as an absolute upper bound on the black hole mass, controls the bias of the underlying microscopic ensemble. The subleading terms represent finite-information corrections to the classical area law.

}

\emailAdd{kehagias@mail.ntua.gr}

\makeatletter
\gdef\@fpheader{}
\makeatother

\begin{document}
\maketitle
\section{Introduction}

The discovery that black holes have entropy \cite{Bekenstein:1973ur} and obey
laws closely analogous to those of ordinary thermodynamics
\cite{Hawking:1975vcx,BekensteinS} has deeply improve our understanding of
gravity. The classical mechanics of stationary black holes can be cast in a
form analogous to the thermodynamic laws, with surface gravity playing the role of
temperature, horizon area that of entropy, and the mass, angular momentum and
charge entering a first-law relation. However, this anlogy with classical GR was purely formal since  the identification of area with
entropy could not be justified, in view of the fact that classical black holes neither radiate nor
equilibrate with an environment.

This picture has changed with the discovery that black holes emit thermal radiation with a
well-defined Hawking temperature~\cite{Hawking:1975vcx}, a Planckian flux of
particles at infinity. The formal analogy between surface gravity and
temperature then became a genuine physical identification, and the
proportionality between entropy and horizon area has acquired a real thermodynamic meaning
(for reviews see e.g.\ \cite{Wald2,Page:2004xp,Witten:2024upt}). Combined with
Bekenstein's earlier arguments, this led to the ``generalized second law'',
which asserts that the sum of ordinary matter entropy outside the black hole and
the horizon entropy never decreases
\cite{Bekenstein:1973ur, Unruh:1982ic,Zurek:1985gd,Frolov:1993fy}.

These developments suggest that black holes are genuine thermodynamic systems
whose entropy and temperature carry direct physical meaning.  On the other hand, they also raise conceptual questions, among which are the microscopic origin of horizon entropy, the failure of black holes to obey the third law, and the need to reconcile the geometric horizon area with the statistical interpretation of entropy as missing information. Such questions have been addressed, for example,
through microstate counting in string and field theory
\cite{Strominger1996,Maldacena:1997de,Strominger1997,Dvali2024,Susskind1994,Susskind1994-2} and
through self-sustained Bose-Einstein condensates (the quantum $N$-portrait)
\cite{Dvali2011,Dvali2015}. Let us note that area-law scaling of  entropy has also been shown  in nongravitational systems at quantum criticality \cite{Dvali:2017nis}.

Even though  the analogy between gravity and thermodynamics is very strong and the thermodynamic laws are satisfied, one  issue remains, namely the third law of thermodynamics. The Nernst
formulation of this law \cite{wilks1961} requires that the entropy $S$
approach zero, or at most a \emph{universal} constant independent of the
macroscopic parameters, as $T\to 0$. For a Schwarzschild black hole,
the temperature decreases as the mass grows while the Bekenstein-Hawking entropy
grows without bound. Indeed,   as $T\to 0$ (formally $M\to\infty$) the entropy diverges
rather than vanishes, which   leads to an apparent violation of the third law
\cite{Wald1,BekensteinS}. 
The latter
persists even when the Hawking temperature vanishes at finite parameters. For example,  an
extremal Kerr black hole has $T=0$ but a finite horizon area, and hence a finite
entropy proportional to the spin (e.g.\ $S=2\pi|J|$). Therefore, it still violates the third law since the entropy does not approaches  a universal
constant independent of $M$ and $J$ at zero temperature.  

There is also the weaker
``unattainability'' formulation of the third law, according to which extremal $T\!=\!0$ states
cannot be reached from non-extremal ones in any finite sequence of processes or
in finite affine time \cite{Bardeen:1973gs,Israel}. Even this has recently been
challenged, as there exist solutions describing gravitational collapse in
general relativity that form an exactly extremal Reissner-Nordstr\"om black hole
in finite time \cite{KU} (see also \cite{Reall1,Crump:2026kgu,Reall2}).

We stress that the violation of the third law need not be a fundamental property of
gravity but it may instead be a feature of GR and its semiclassical implementation only, which remains  
to be resolved in a more complete theory at low temperatures (large masses).

The familiar case of black-body radiation is instructive. Using only the
high-energy behaviour of a photon gas with an unbounded spectrum and the
corresponding asymptotic density of states, the microcanonical entropy grows
only logarithmically, $S(E)\sim \ln E$. Taken at face value this behavior 
violates the Nernst theorem, since the entropy  $S$ diverges for low  energies ($E\to 0$). Yet, as Planck showed, this
conclusion is wrong since once the full spectrum is included,  the canonical
entropy $S(T)$ satisfies the Nernst third law, while $S\sim \ln E$ survives
only as a high-temperature asymptotic form. In fact, it is the true ground state and full
quantum structure of the spectrum, not the leading high-energy behavior, that
determine the low-temperature limit.

Black hole entropy  may be similar, with the Bekenstein-Hawking formula a
semiclassical approximation valid in a particular regime.  It is therefore
natural to regard the Bekenstein-Hawking expression $S\sim M^2$ as the leading term of a
more complete entropy function, just as the naive $S\sim\ln E$ for black-body
radiation is superseded by the full Planck formula, restoring compatibility with
the Nernst theorem and the third law.

This is what we are proposing in this work. Namely, we impose the Nernst third law as an additional requirement on
black holes. Guided by the black-body analogy, we first reconstruct the entropy
of a photon gas from its high-energy behavior once the third law is enforced,
and then apply the same strategy to gravity, deriving a modified gravitational
entropy that satisfies the Nernst law while reducing to the Bekenstein-Hawking
area law in the appropriate regime. We thus construct a one-parameter deformation of the Schwarzschild entropy,
controlled by the universal mass scale $M_0$, which reproduces the
Bekenstein--Hawking area law for $M/M_0\ll1$. By construction, the resulting
entropy satisfies the third law, approaching a finite universal constant in
the zero-temperature limit. Remarkably, the same function is exactly the
relative entropy of a biased Bernoulli product measure with respect to its
unbiased reference measure. It therefore admits an information-theoretic
representation as an entropy deficit,\footnote{An
entropy deficit is often called a negentropy~\cite{Schrodinger1944,Brillouin1956}.}
relative to the maximally mixed ensemble.

\section{Black Hole Entropy}

The black hole entropy, given by the Bekenstein-Hawking formula, neither
vanishes nor approaches a ``universal'' constant independent of the intensive
thermodynamic parameters in the limit of zero temperature, as required by the
Nernst formulation of the third law of thermodynamics~\cite{wilks1961}. However,  although the applicability of the third law to black holes has been questioned
in the literature~\cite{Wald1,Wald2},   it remains a natural and
interesting challenge to search for a black hole entropy formula for which
the third law is also obeyed.
 In particular, we will require here that the black hole entropy
 satisfies
\begin{equation}
\lim_{T \to 0} S = S_0,
\label{s}
\end{equation}
where $S_0$ is a \emph{universal constant} (the residual entropy), independent
of the intensive thermodynamic parameters of the system. 
In other words, we will assume that the Nernst's third law is valid for ordinary, neutral, non–rotating
Schwarzschild black holes.  Rotating and/or charged black holes will be considered elsewhere.
Equivalently, defining the  heat capacity  as 
\begin{eqnarray}
C_V=T\frac{dS}{dT},
\end{eqnarray}
where the temperature is determined by
\begin{eqnarray}
\frac{1}{T}=\frac{dS}{dE},
\end{eqnarray}
 the Nernst condition
\eqref{s} for the regular entropy functions considered here implies that the heat capacity vanishes in the zero-temperature
limit
\begin{eqnarray}
\lim_{T\to0}C_V=0.
\label{Cv}
\end{eqnarray}

Let us
 recall that the Bekenstein-Hawking entropy (B-H) for a Schwarzschild black hole of mass $M$ is 
\begin{eqnarray}
  S_{ B\!-\!H}=4\pi M^2, \label{entropy}
  \end{eqnarray}  
($G_N=1$) and its corresponding temperature is
\begin{eqnarray}
T=\left(\frac{dS_{\BH}}{dM}\right)^{-1}=\frac{1}{8\pi M}. \label{tem} 
\end{eqnarray}
The zero-temperature limit $T\to 0$ corresponds to $M\to \infty$,  where however, the entropy diverges $S\to \infty$. Similarly, the  heat capacity of a black hole turns out to be
\begin{eqnarray}
 C_V=-\frac{1}{8\pi} \frac{1}{T^2}, \label{cvv}
 \end{eqnarray} 
 which also diverges as $T\to 0$.
 Neither the entropy, nor the specific heat satisfies Eqs. (\ref{s}) and (\ref{Cv}), respectively. Therefore, black holes do not obey the third law of thermodynamics.
 
 Our target is to search for an entropy formula that satisfied the third law of thermodynamics and reduces to the usual Bekenstein-Hawking formula at appropriate limit. 
 A possible guide towards this target is the black body radiation analog, which we will now examine. 
 
 \vskip.1in
 \noindent 
 \subsection{The Blackbody case}

 It is known that, in Planck's thermodynamic treatment of blackbody
radiation, the high-energy entropy has the form
 \begin{eqnarray}
   S=\ln E.  \label{ss1}
   \end{eqnarray}  
The temperature is 
$T=E$, and therefore, the third law of thermodynamics is  violated  since the entropy $S=\ln T$ diverges as $T\to 0$. However, using quantum statistical mechanics, the correct expression for the entropy turns out to be given by the Planck formula
\begin{eqnarray}
S_{\rm rad}=\frac{1}{2}\left(1+\frac{E}{E_0}\right)\ln \left(\frac{E}{E_0}+1\right)
-\frac{1}{2}\left(\frac{E}{E_0}-1\right)\ln \left(\frac{E}{E_0}-1\right)-\ln 2, \label{ss}
\end{eqnarray}
where $E_0=\hbar \omega_0/2$ is the ground state energy of a harmonic oscillator.\footnote{Note that the expression in Eq. (\ref{ss}) for the entropy  includes also  the zero-point energy of the corresponding mode. The transformation $E\to E+E_0$ brings Eq. (\ref{ss}) to the form written by Planck in first place. Later, he introduced the notion of zero-point energy, which is an interesting story \cite{Mehra}.   } 
From the above expression, we find that 
\begin{eqnarray}
\frac{1}{T}=\frac{1}{2E_0}\ln\left(\frac{E+E_0}{E-E_0}\right).
\end{eqnarray}
Therefore, the limit $T\to 0$ corresponds to  $E\to E_0$ where from Eq.(\ref{ss}) $S\to 0$, in accordance with the third law of thermodynamics.  In other words, the entropy Eq. (\ref{ss})  satisfies also the third law of thermodynamics, and  it is only  its high energy approximation that violates it. Of course,  Eq. (\ref{ss}) is derived by employing quantum statistical mechanics. 
So the question one could ask  is the following: 

\medskip
\medskip
\noindent
{\it Is it possible to derive Planck's formula (Eq.~\ref{ss}) 
by imposing only the third law of thermodynamics and the 
high-energy asymptotic behavior of entropy?}

\medskip
\medskip 
\noindent 
The answer is  affirmative under some mild extra assumption which I'll describe below. 
 Since the entropy in Eq.(\ref{ss1}) diverges as $E\to 0$, we will assume that there exist a minimum energy, $E_{min}=E_0$, where both the entropy and the temperature  vanishes, i.e., 
\begin{eqnarray}
T(E_0)=0, ~~~~~~S(E_0)=0. 
\end{eqnarray}
 At this point $dS/dE$ should diverge, since it is the inverse of the temperature. This divergence could be either logarithmic or a pole singularity so that the second derivative $d^2S/d^2E$ should have a  pole. Therefore, we expect a behavior of the form
\begin{eqnarray}
 \frac{d^2S_{\rm rad}}{dE^2}=-\frac{a_n(E) }{(E-E_0)^n}+\cdots 
- \frac{a_2(E)}{(E-E_0)^2}-\frac{a_1(E)}{(E-E_0)} \, ,  \label{se}
 \end{eqnarray}
where  the functions $a_i(E)$ for  $(i=1,\cdots,n)$ are analytic function for $E>0$ and scale as 
\begin{eqnarray}
    a_i(E)\sim E^{i-2} \quad \mbox{for}
    \quad E\gg E_0,
\end{eqnarray}
in order  to get  Eq. (\ref{ss1}) for the entropy.
%
 Notice that we have not included   analytic terms  of $E$ in Eq. (\ref{se}) since such terms  will dominate the logarithmic growth of the entropy at high energies.  For example,  a term proportional to the  energy  $E$ in Eq.(\ref{se}) would lead to the behavior $S\sim E^3$, which does not match the desired one $S= \ln E$ of Eq. (\ref{ss1}).  We should also demand, according to the third law of thermodynamics, that the entropy should remain finite at $T=0$, i.e for $E=E_0$. It is easy then to see that all terms proportional to $a_n$ for $n\neq 1$, lead to a divergence in the entropy at $E=E_0$. The only term in     Eq. (\ref{se}) which gives  finite entropy at $E=E_0$ arises from the simple pole term, proportional to $a_1(E)$, so that 
 \begin{eqnarray}
 \frac{d^2S_{\rm rad}}{dE^2}=-\frac{a_1(E)}{(E-E_0)} .  \label{see1}
 \end{eqnarray}
 Since $a_1$  scales like $1/E$, its simpler form is 
\begin{eqnarray}
 a_1(E)=\frac{c}{E+\epsilon}, 
 \end{eqnarray} 
 where $c$ and $\epsilon>0$ are positive constants. 
 In others words, the second derivative of the entropy has  the form 
 \begin{eqnarray}
 \frac{d^2S_{\rm rad}}{dE^2}=-\frac{c}{(E-E_0)(E+\epsilon)} \, ,  \label{se2}
 \end{eqnarray}
 in order to have finite entropy in the $T\to 0$ limit and the correct logarithmic behaviour of Eq. (\ref{ss1}) at large energies.\footnote{Actually, this is the form Planck assumed for the second derivative of the entropy after inclucing the zero-point energy.}  
  Then, integrating twice Eq. (\ref{se2}) we get 
 \begin{eqnarray}
 S_{\rm rad}=\frac{c}{E_0+\epsilon}\Big[\big(E+\epsilon\big)\ln\big(E+\epsilon\big)
 -\big(E-E_0\big)\ln\big(E-E_0\big)\Big] +\alpha E+S_0, \label{sss}
 \end{eqnarray}
 where $\alpha$ and $S_0$ are integration constants. At $E=E_0$, the entropy  remains finite as it should be, according to the third law of thermodynamics.
 Therefore, we take $a_n=0$ for all $n\neq 0$, whereas, the coefficients $c$, $\epsilon$ and $\alpha$ are determined by the high energy behaviour of Eq.(\ref{ss1}). Indeed, expanding Eq. (\ref{sss}) 
 for large $E$, we get
 \begin{eqnarray}
 S_{\rm rad}\approx c \ln E+\alpha E+\frac{c}{E_0+\epsilon} \left(
 \frac{\epsilon^2-E_0^2}{E}-\frac{\epsilon^3+E_0^3}{E^2}+\cdots\right)+S_0.
 \label{sss1}
 \end{eqnarray}
 The values of $c$ and $\alpha$ are determined by the high energy behavior of the entropy in Eq. (\ref{ss1}), whereas the value of $\epsilon$ is determined from the vanishing of the $E^{-1}$ term in  Eq.(\ref{sss1}) since retaining such a term  will correspond to a constant shift in the energy. Since the zero of energy has not  been fixed, we may choose $\epsilon=E_0$ so that the large-energy relation becomes $T\simeq E$.
 These conditions determine the value of the constants $c,~\alpha$ and $\epsilon$ to be  
 \begin{eqnarray}
 c=1, ~~~~~~~\alpha=0, ~~~~~~~\epsilon=E_0. 
 \end{eqnarray}
 Hence, the final expression for the entropy after choosing $S_0$ so that $S_{rad}\to 0$ for $E\to E_0$ is  
 \begin{eqnarray}
 S_{\rm rad}=\frac{1}{2E_0}\Big[\big(E+E_0\big)\ln\big(E+E_0\big)
 -\big(E-E_0\big)\ln\big(E-E_0\big)\Big] -\ln 2. \label{s3}
 \end{eqnarray}
This is identical to the expression for the entropy of black body radiation given in Eq. (\ref{ss}).  
Summarizing, the blackbody entropy is obtained upon introducing a minimum energy 
$E_0$ such that $T(E_0)=0$, 
 thereby ensuring vanishing  entropy ($S(E_0)=0$) consistent with the third law of thermodynamics.

 
\vskip.2in
\noindent
\subsection{Back to Black Holes}
\begin{figure}[t!]
	\centering
	\includegraphics[width=.45 \linewidth]{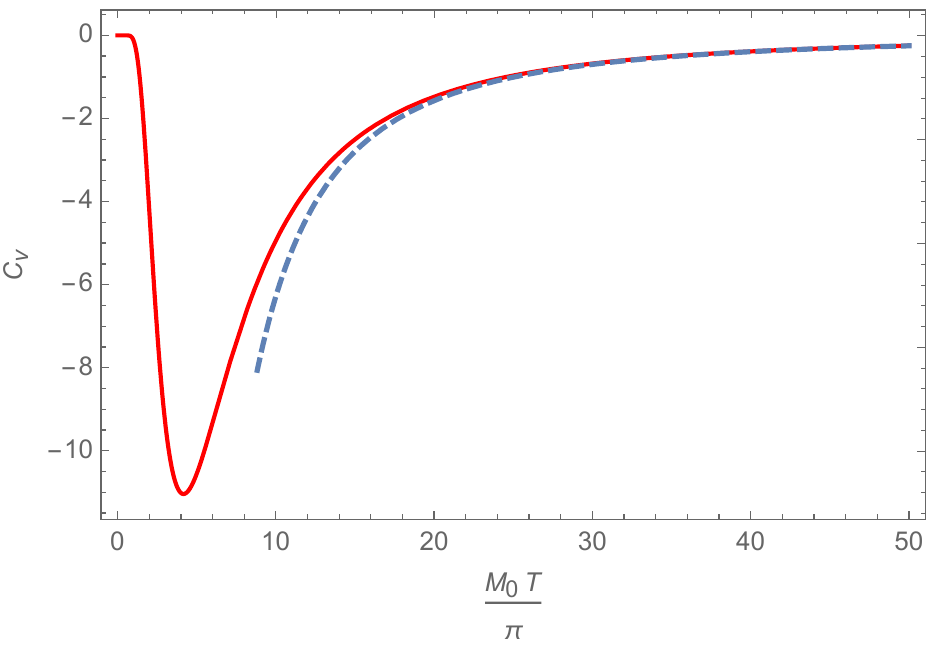}
	\includegraphics[width=.45 \linewidth]{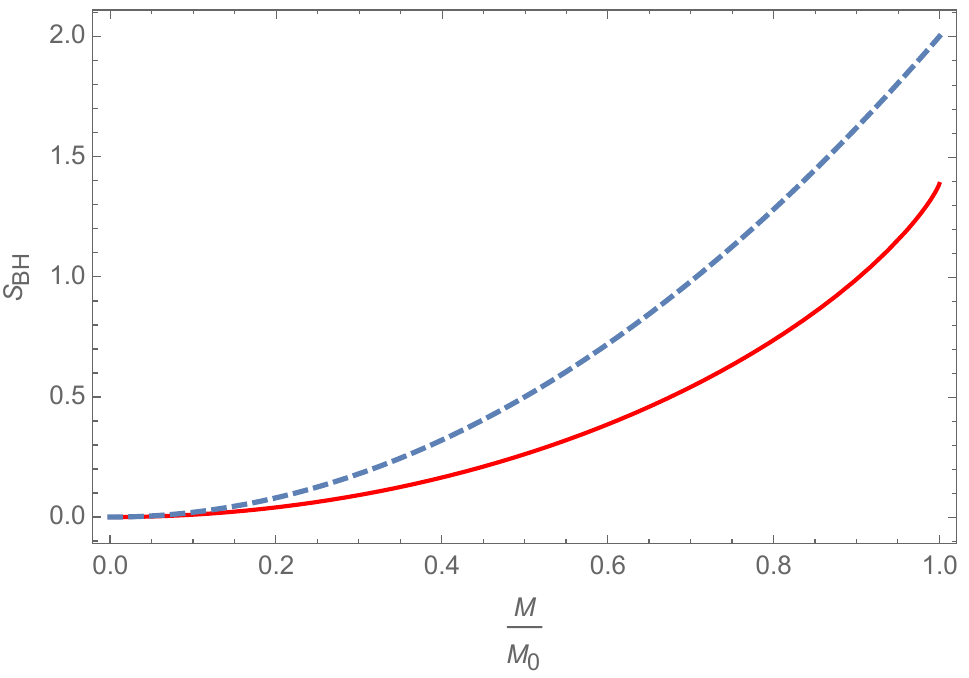}
	\caption{\it 
	{\bf Left:}
	The heat capacity  described by Eq. (\ref{BHc}) is the red curve whereas the dashed blue curve describes Eq. (\ref{cvv}). We see that the two curves coincide for high temperatures, but only the red one satisfies the third law of thermodynamics. 
	{\bf Right:} The black hole entropy in units of $8\pi M_0^2$ for Eq. (\ref{sm3}) is indicated with red color whereas, the Bekenstein-Hawking entropy is the dashed blue curve. They coincide only at high temperatures.
	}
	\label{fig1}
\end{figure}

Let us now return to the black hole case and follow  the same procedure as above, aiming to  an  entropy formula similarly consistent with the   the third law of thermodynamics.    
  Here, since $T\to 0$ corresponds to $M\to \infty$, we will assume  that there is a maximum value 
\begin{eqnarray}
  M_{\rm max}=M_0, \qquad M\leq M_0, 
  \end{eqnarray}  
  for the black hole mass $M$,
  instead of a minimum value as in the black body case, where the temperature is zero and the entropy is constant,   
\begin{eqnarray}
  T(M_0)=0, ~~~~\mbox{and}~~~~S_{\bh}(M_0)=\mbox{finite}. 
  \label{sfinite}
  \end{eqnarray}  
The scale $M_0$ denotes the maximum mass a black hole can attain, which could, in principle, be related to the total mass of the Universe. In other words, $M_0$ represents the mass of a black hole that would result from the complete gravitational collapse of the entire Universe\footnote{Since the mass of the Universe is epoch-dependent, it is not an ideal quantity to tie to a fixed scale. In fact, as we will show in the accompanying paper~\cite{kehagias-2}, the scale $M_0$ is actually related to the cosmological constant.}.
Thus, given Eq. \eqref{sfinite} and following the discussion of the blackbody radiation, $dS_\bh/dM$ should have a logarithmic or a pole singularity, so that  $d^2S_\bh/dM^2$ should have a pole singularity at $M=M_0$,  while its behavior away from the endpoint must allow the
Bekenstein--Hawking limit to be recovered for $M\ll M_0$. 
It is easy to verify again that higher order poles lead to  divergences in the entropy at $M=M_0$ so that we ignore them.  These conditions
do not uniquely determine the entropy. As a minimal rational ansatz,
analogous to the blackbody case, we take
\begin{eqnarray}
 \frac{d^2S_{\bh}}{dM^2}=\frac{b_1}{(M_0-M)(M+m)} \, ,  \qquad M\leq M_0 \label{sm}
 \end{eqnarray} 
 where $m>0$ and $b_1>0$ are constants to be fixed by matching the
small-mass expansion to the Bekenstein--Hawking entropy.  
Moreover, since in general 
\begin{eqnarray}
\frac{d^2S_\bh}{dM^2}=-\frac{1}{T^2C_V}, 
\end{eqnarray}
and since $C_V<0$ for black holes, the entropy is a convex function here and therefore $b_1>0$. 
Integrating  twice Eq. (\ref{sm}), we find that the entropy is given by
 \begin{eqnarray}
S_\bh=   \frac{b_1}{M_0+m}\Big[\big(M_0-M\big)\ln \big(M_0-M\big)+\big(M+m\big)\ln\big(M+m\big)\Big] +S_0. \label{sm1}
   \end{eqnarray}  
  Expanding Eq.(\ref{sm1}) for $M\ll M_0$, we get 
  \begin{eqnarray}
  S_{\bh}&\approx&\frac{b_1}{M_0+m}\Big(M_0 \ln M_0+m\ln m\Big) +S_0-\nonumber \\
  &&- \frac{b_1}{M_0+m}\ln\left(\frac{M_0}{m}\right)\, M+
  \frac{b_1}{2mM_0}M^2-\frac{b_1\Big(M_0-m\big)}{6m^2M_0^2}M^3+\cdots \, . \label{sm2}
  \end{eqnarray}
  Therefore, in order  the leading term of Eq. (\ref{sm2}) to agree with 
  the Bekenstein-Hawking entropy Eq. (\ref{entropy}), i.e., 
  \begin{eqnarray}
      S_{\bh}\approx S_{\rm B-H}=4\pi M^2,\qquad \mbox{for}\qquad M\ll M_0
  \end{eqnarray}
  we should have 
  \begin{eqnarray}
  m=M_0, \qquad b_1= 8\pi M_0^2, \qquad S_0=-8\pi M_0^2\,\ln M_0,
  \end{eqnarray}
  In this case, the expression for the black hole entropy turns out to be
  \setlength{\fboxsep}{12pt}
  \begin{eqnarray}
  \boxed{
S_{\bh}=   4\pi M_0^2\bigg[\left(1-\dfrac{M}{M_0}\right)\ln \left(1-\dfrac{M}{M_0}\right)+\left(1+\dfrac{M}{M_0}\right)\ln\left(1+\dfrac{M}{M_0}\right)\bigg],}
\label{sm3}
   \end{eqnarray} 
  and it is plotted with red color in the right panel of Fig. (\ref{fig1}). 
  The black hole temperature is then 
  \begin{eqnarray}
  \boxed{
  \frac{1}{T}=4\pi M_0\ln\left(\dfrac{M_0+M}{M_0-M}\right), \label{temp}
  }
  \end{eqnarray}
  which for small black hole masses $M\ll M_0$ is approximately 
  \begin{eqnarray}
  \frac{1}{T}\approx 8\pi M, 
  \end{eqnarray}
  as expected. Similarly, the heat capacity is 
  \begin{eqnarray}
 \boxed{
   C_V=-\frac{1}{8\pi T^2}\,{\rm sech}^2\left(\dfrac{1}{8\pi M_0T}\right).  \label{BHc}
   }
   \end{eqnarray} 
 from  where the expression Eq. (\ref{cvv}) is recovered for small black hole masses (high temperature). However, for low temperature we get 
  \begin{eqnarray}
  C_V\to 0 ~~~~~\mbox{for} ~~~~~T\to 0,
  \end{eqnarray}
  in accordance with the third law of thermodynamics. 
The behavior of the heat capacity is illustrated in  the left panel of Fig. (\ref{fig1}).

It is clear from the right panel in Fig. (\ref{fig1}) that the entropy is a convex function of the  energy. This is actually what differentiate our formula for black hole entropy  to the one of the black body. In fact, the general expression of the second derivative of the 
entropy which  does not lead to divergences in the entropy is 
\begin{eqnarray}
 \frac{d^2S}{dE^2}=-\frac{a_1}{(E-E_0)(E+E_0)}, ~~~~~a_1>0.
 \end{eqnarray} 
There are two possibilities now: 1) the entropy is a concave function of the energy and therefore  $E_0$ represents a minimum energy (ground state), or, 2) the entropy is  convex  in which case, $E_0$ represents  a maximum energy. The case (1) corresponds to the black body and case (2) corresponds to the black hole, or at least to a system with similar thermodynamic properties. It is clearly thermodynamically unstable as any system with convex entropy.  

One might be tempted to identify a Schottky-like anomaly in the heat-capacity behavior described by Eq.~\eqref{BHc} and shown in Fig.~\eqref{fig1}. However, the curve in Fig.~\eqref{fig1} does not display a conventional Schottky anomaly. Instead, it exhibits an anomalous dip into negative  heat capacity, pointing to the nonstandard thermodynamic behavior of gravity rather than the usual Schottky response of a gapped two-level system. However, there is a relation to a two-lever system, which will be  discussed later. 

As one can see from Fig.\ref{fig1},  the entropy \eqref{sm3} exhibits a maximum for $M/M_0=1$. 
In order to determine the  exact value of the maximum, we take the limit $M\to M_0$ in Eq. (\ref{sm3}), from where we find that 
\begin{eqnarray}
S_{\rm max}=8\pi M_0^2 \ln 2=\ln2^{8\pi M_0^2}. \label{smax}
\end{eqnarray}
By writing this entropy as 
\begin{eqnarray}
 S_{\rm max}=\ln 2^N,
\label{S_N}
 \end{eqnarray} 
 we get, after introducing back the reduced Planck mass $M_P$, that $N$ is given by  
 \begin{eqnarray}
 N= \frac{M_0^2}{M_P^2}. \label{N}
 \end{eqnarray}
 It is clear from the form of Eq. (\ref{S_N}) that the quantity $N$ is simply the entropy measured in bits
$
N=\frac{S_{\max}}{\ln 2}=\log_{2}\Omega_{\max},
$
where $\Omega_{\max}$ is the number of accessible microstates at maximal entropy. In this sense $N$ represents the maximum information capacity (maximum number of bits) under the assumed macroscopic constraints for a black hole of mass $M=M_0$. 
  Therefore, the black hole entropy can be written in terms of $N$ as  
 \begin{align}
     S_{\bh}= &  \frac{N}{2}\bigg[\left(1-\frac{M}{M_0}\right)\ln \left(1-\frac{M}{M_0}\right)+\left(1+\frac{M}{M_0}\right)\ln\left(1+\frac{M}{M_0}\right)\bigg].\label{sm5}
 \end{align}
 Expanding  $S_{\bh}$ in  powers of $M/M_0<1$, we get, after using Eq.(\ref{N}),
 \begin{align}
     S_{bh}=&\frac{M^2}{2 M_P^2}
     \left(1+\frac{1}{ N} \frac{M^2}{6 M_P^2}+\frac{1}{ N^2}\frac{M^4}{15 M_P^4}+\cdots
     \right)\nonumber \\
     =&\frac{x}{2} \sum_{n=0} \frac{1}{N^n}\frac{x^{n}}{(2n+1)(n+1)}, \qquad  x=\frac{M^2}{M_P^2}. \label{SN}
 \end{align}
 Hence, the anticipated entropy formula (\ref{sm3}), or (\ref{sm5}), decomposes into a leading contribution of
order $N^0$ plus an infinite tower of subleading terms suppressed by powers
of $1/N$.
In the limit  $N\to\infty$ with $x$ 
fixed, only the first term in
Eq.~\eqref{SN} survives,
\begin{equation}
  S_{\rm bh}(M)
  \xrightarrow[N\to\infty]{}\frac{x}{2}=
  \frac{M^2}{2M_P^2},
\end{equation}
so that the entropy reduces to the usual Bekenstein-Hawking entropy.  This is the analogue of the
thermodynamic limit in standard statistical mechanics, in which the entropy
becomes an extensive, smooth function of the macroscopic variables and
fluctuations are negligible.  In our model, this leading term is precisely
the anticipated macroscopic entropy used in Eq.~(3.8) and Eq.~(3.17).

Notice also that we can write the black hole entropy (\ref{SN})  in an $1/N$ expansion as a positive series 
\begin{eqnarray}
    S_{\bh}&=&\frac{A}{4}\left(1+\frac{1}{N} \frac{A}{12}+ \frac{1}{N^2}\frac{A^2}{30}+\cdots\right)\nonumber \\
    &=&\frac{A}{4}\sum_{k=0}\frac{1}{N^k} \frac{A^k}{2^k (2k+1)(k+1)}, \label{SA}
\end{eqnarray}
 where $A$ is the horizon area in units of $\ell_P^2$. Hence, the black hole entropy is a increasing function of the horizon area whenever an area theorem, as in GR, holds.

Notice that, since the entropy at $T=0$ is $S_{\rm max}$, it should be a universal constant according to the third law. Therefore, $M_0$ should be a new universal constant, describing the maximum mass, or maximum Schwarzschild radius any black hole may have.

\section{Microcanonical Ensemble for Black Hole Entropy}

In Eq.~(\ref{sm3}) we have introduced the black-hole entropy $S_{\rm bh}$ as a
purely macroscopic quantity, defined in terms of the black hole mass $M$ and a universal
mass scale $M_0$. In this section we would like to go beyond this
macroscopic definition and ask what kind of microscopic degrees of freedom, if any,
could give rise to such an entropy. Our aim is to obtain a simple
statistical interpretation of $S_{\rm bh}$, in which its dependence on $M$
can be traced back  in some underlying microscopic variables.

A convenient way to make this structure explicit is to recast $S_{\rm bh}$
in a form that resembles the entropy of a statistical system. Using
Eq.~(\ref{sm3}), one can easily show that $S_{\rm bh}$ may be written as
\begin{eqnarray}
  &&  S_{\bh}=N \ln 2+ N\Bigg\{ \frac{1}{2}\left(1+ \frac{M}{M_0}\right) \ln \left[\frac{1}{2}\left(1+\frac{M}{M_0}\right)\right]+\frac{1}{2}\left(1-\frac{M}{M_0}\right) \ln \left[\frac{1}{2}\left(1-\frac{M}{M_0}\right)\right]\Bigg\}.\nonumber \\
    && \label{micro}
\end{eqnarray}
This expression makes it manifest as we will see in the sequence, that $S_{\rm bh}$ is associated with the
entropy of $N$ independent Bernoulli trials \cite{GrimmettStirzaker2001ProbabilityRandom} with a mass-dependent bias
controlled by the ratio $M/M_0$. Equation~(\ref{micro})  will
serve as the starting point for the microscopic interpretation
developed in the remainder of this section. 

Starting from Eq.~(\ref{micro}), it is convenient to
introduce the shorthand notation
\begin{equation}
  p = \frac{1}{2}\!\left(1+\frac{M}{M_0}\right),
\qquad
  1-p = \frac{1}{2}\!\left(1-\frac{M}{M_0}\right),
\label{eq:def-p}
\end{equation}
so that the two coefficients appearing in Eq.~(\ref{micro}) are
simply the probabilities $p$ and $(1-p)$ of a Bernoulli distribution.  In
terms of this bias parameter the entropy can be written in the compact form
\begin{equation}
  S_{\rm bh}
  = N \ln 2
    + N\Big[\,p \ln p + (1-p)\ln (1-p)\,\Big]
\label{eq:Sbh-compact}
\end{equation}
or, equivalently,
\begin{equation}
  S_{\rm bh} = \ln \!\Big[ 2\,p^{\,p}(1-p)^{\,1-p} \Big]^{\!N}.
\end{equation}
This expression makes explicit that all the dependence on the black-hole
mass $M$ is encoded in the single parameter $p(M)$, while $N$ controls the
overall size of the underlying microscopic system.

It is useful to rewrite Eq.~\eqref{eq:Sbh-compact} in a way that highlights
its statistical interpretation.  To this end, note that the term in brackets
can be expressed in terms of the Shannon entropy of a Bernoulli variable with
bias $p$ \cite{CoverThomas2006},
\begin{equation}
  H(p) = -\,p \ln p - (1-p)\ln (1-p),
\label{eq:Shannon-Bernoulli}
\end{equation}
so that Eq.~\eqref{eq:Sbh-compact} becomes
\begin{equation}
  S_{\rm bh}
  = N\ln 2 - N H(p).
\end{equation}
Exponentiating both sides, we obtain
\begin{equation}
  e^{S_{\rm bh}}
  = \frac{2^N}{e^{N H(p)}}.
\end{equation}
This suggests introducing the notation
\begin{equation}
  W_{\rm tot} = 2^N,
  \qquad
  W_{\rm vis} = e^{N H(p)}=2^{N H_2(p)},
\label{eq:Wtot-Wvis-def}
\end{equation}
where 
\begin{eqnarray}
    H_2=-p\log_2 p-(1-p)\log_2 (1-p), \label{H2}
\end{eqnarray}
is the Shannon entropy (\ref{eq:Shannon-Bernoulli}) in bits. Therefore, the black hole entropy $S_{\bh}$ is 
 written as
\begin{equation}
  S_{\rm bh}
  = \ln \!\left(\frac{W_{\rm tot}}{W_{\rm vis}}\right).
\label{eq:Sbh-ratio}
\end{equation}

The quantities $W_{\rm tot}$ and $W_{\rm vis}$ admit a natural microscopic
interpretation.  Consider $N$ binary degrees of freedom (for instance,
Ising spins, bits, or individual steps of a one-dimensional random walk).
If each degree of freedom is completely unbiased, there are
$
  W_{\rm tot} = 2^N
$
distinct microstates, all a priori accessible.  On the other hand, if the
system is constrained to reproduce a macroscopic bias $p$ (for example, a
fixed average magnetization, or a fixed fraction $p$ of ``up'' spins), then
only a subset of all $2^N$ configurations is compatible with that macroscopic
constraint.  In the language of information theory, the number of
typical configurations of $N$ biased Bernoulli variables is of order
$2^{N H_2(p)}$, which we have denoted by $W_{\rm vis}$ in
Eq.~\eqref{eq:Wtot-Wvis-def}.

Thus, Eq.~\eqref{eq:Sbh-ratio} shows that the black hole entropy can be
viewed as the logarithm of the ratio between the total number of microscopic
states of an unbiased $N$-bit system and the number of states that remain
``visible'' once we impose the macroscopic bias $p(M)$ associated with a
black hole of mass $M$.  In this sense, $S_{\rm bh}$ measures how strongly
the microscopic ensemble is constrained by the presence of the black hole. In particular, 
the larger $S_{\rm bh}$, the larger the reduction of the accessible
configuration space when we condition on the macroscopic parameter $M$.

The representation \eqref{eq:Sbh-ratio} admits also a particularly transparent
interpretation in terms of visible and hidden microstates.
Consider again an ensemble of $N$ Bernoulli variables $X_i \in \{0,1\}$.
A microscopic configuration (or ``microstate'') is specified by a complete
string
\begin{equation}
  \vec X = (X_1,X_2,\dots,X_N),
\end{equation}
so that, in the absence of any constraint, the total number of distinct
microstates is
\begin{equation}
  W_{\rm tot} = 2^N.
\end{equation}
This corresponds to the reference ensemble of $N$ unbiased bits, in which
all $2^N$ configurations are a priori equally likely and accessible.

Now impose the macroscopic constraint that the ensemble has a bias $p$
towards one of the two outcomes, as in Eq.~\eqref{eq:def-p}.  From the
point of view of an exterior observer who only probes the aggregate
properties of the system, the relevant configurations are those that
realize this macroscopic bias.  Information theory tells us that, in the
limit of large $N$, the number of such typical configurations is
exponentially smaller than $2^N$ and is given by
\begin{equation}
  W_{\rm vis} = 2^{N H_2(p)},
\end{equation}
where $H_2(p)$ is the Shannon entropy in bits of the Bernoulli distribution with
parameter $p$, given in Eq. (\ref{H2}). 
We therefore interpret $W_{\rm vis}$ as the number of microstates that are
effectively ``visible'' or distinguishable to an exterior observer who only
has access to the biased statistics encoded in $p(M)$.
 With these definitions, one finds
\begin{equation}
S_{\rm bh} = N\ln 2\Big[1-H_2(p)\Big]=\ln\left(\frac{W_{\rm tot}}{W_{\rm vis}}\right).
\end{equation}
Thus, $S_{\rm bh}$ remains the thermodynamic black hole entropy, but in the
present Bernoulli representation it takes the form of an entropy deficit
relative to the maximally mixed ensemble. Notice that
$W_{\rm tot}/W_{\rm vis}$ need not be an integer and should therefore not be
interpreted as an exact number of hidden microstates.

\begin {comment}

It is often useful to emphasize this point by rewriting $S_{\rm bh}$ again
in terms of the Shannon entropy of the visible Bernoulli ensemble.  From
Eq.~\eqref{eq:Sbh-compact} and the definition of $H(p)$ we have
\begin{equation}
  S_{\rm bh}
  = N\ln 2 - N H(p).
\end{equation}
This relation shows that $S_{\rm bh}$ is, strictly speaking, not the entropy
of  degrees of freedom themselves, but an \emph{entropy
deficit} \cite{Brillouin1953,KullbackLeibler1951,Landsberg1984,CoverThomas2006}.  It measures, on a logarithmic scale, by how much the coarse‐grained, biased description of the system (with entropy $N H(p)$) falls
short of the maximal entropy $N\ln 2$ available in the unbiased reference
ensemble \cite{Brillouin1956}.  Equivalently, $S_{\rm bh}$ quantifies how much information about
the full microscopic configuration is hidden from the exterior observer.

To make this even more concrete, recall that for an unbiased Bernoulli
distribution ($p=\tfrac{1}{2}$), corresponding in our parametrization to $M=0$,
the Shannon entropy is maximized,
\begin{equation}
  H_{\rm uniform} = H\!\left(\tfrac{1}{2}\right) = \ln 2.
\end{equation}
In this case 
\begin{eqnarray}
    N H\!\left(\tfrac{1}{2}\right) = N\ln 2, \qquad\Longrightarrow \qquad
    S_{\rm bh}=0
\end{eqnarray}
so that 
no information is hidden, and the coarse-grained
description is already maximally entropic.  As soon as $p$ departs from
$1/2$ (i.e.\ as soon as a black hole with $M\neq 0$ is present), the
visible entropy $N H(p)$ decreases, $W_{\rm vis}$ becomes a strict subset of
$W_{\rm tot}$, and $S_{\rm bh} = \ln W_{\rm hid}$ becomes positive.  The
larger the deviation of $p$ from $1/2$, the larger the entropy deficit and
hence the larger the number of hidden microstates required to account for
the macroscopic configuration.  In this way, the formula for $S_{\rm bh}$
naturally acquires the interpretation of an entropy of missing information
associated with degrees of freedom that are effectively ``behind the
horizon'' of the exterior observer.

\end{comment}

\section{Black Hole Entropy as Kullback-Leibler  divergence }

The interpretation of $S_{\rm bh}$ 
as the logarithm
of a hidden multiplicity becomes even more transparent when we rewrite it in
terms of the Kullback-Leibler (KL) divergence \cite{KullbackLeibler1951}.  Recall that for two discrete
probability distributions $P=\{P_i\}$ and $Q=\{Q_i\}$ on the same sample
space, the KL divergence of $P$ relative to $Q$ is defined by
\begin{equation}
  D_{\rm KL}(P\Vert Q)
  = \sum_i P_i \ln \frac{P_i}{Q_i}.
\label{eq:KL-def}
\end{equation}
From an information-theoretic perspective, $D_{\rm KL}(P\Vert Q)$ measures
the additional average number of nats (or bits, if the logarithm is taken in
base~2) required to encode samples drawn from $P$ when one uses a code
optimized for $Q$ rather than for $P$ itself.  Equivalently, it quantifies
how much information is gained, on average, when one updates a prior $Q$ to a
posterior $P$.  The KL divergence is non-negative and vanishes if and only
if $P=Q$, so it behaves as an asymmetric ``distance'' between probability
distributions.

The KL divergence is  closely related to standard entropy measures.  The
Shannon entropy of $P$ is
\begin{equation}
  H(P) = -\sum_i P_i \ln P_i,
\end{equation}
whereas the cross-entropy of $P$ relative to $Q$ is
\begin{equation}
  H(P,Q) = -\sum_i P_i \ln Q_i.
\end{equation}
In terms of these quantities,
\begin{equation}
  D_{\rm KL}(P\Vert Q) = H(P,Q) - H(P),
\end{equation}
so the KL divergence is precisely the gap between the entropy of $P$ and the
average code length obtained when encoding $P$ with a code designed for $Q$.
In nonequilibrium statistical mechanics, this same quantity appears as a
measure of excess free energy. For a system at temperature $T$ with a
nonequilibrium distribution $P$ and an equilibrium distribution $Q$, the
difference between the corresponding free energies can be written as
\begin{equation}
  F(P) - F(Q) = k_{\rm B} T\, D_{\rm KL}(P\Vert Q),
\end{equation}
so that $D_{\rm KL}(P\Vert Q)$ directly measures the amount of free energy
(or irreversible work) available as the system relaxes from $P$ to $Q$.

In our case the relevant distributions live on a two-point space.  We
consider a binary outcome ($i=\pm$) with probabilities
\begin{equation}
  p_\pm = p_\pm(M)
  = \frac{1}{2}\!\left(1 \pm \frac{M}{M_0}\right),
\end{equation}
which encode the mass dependence of the black hole through the bias $M/M_0$.
We take as a reference distribution the unbiased Bernoulli law
\begin{equation}
  q_+ = q_- = \frac{1}{2},
\end{equation}
corresponding to the maximally entropic situation $M=0$ discussed above.
The KL divergence of the biased distribution $p=\{p_+,p_-\}$ relative to this
uniform reference is then
\begin{equation}
  D_{\rm KL}\!\left(p\Big\Vert\frac{1}{2}\right)
  = \sum_{i=\pm} p_i \ln \frac{p_i}{1/2}
  = p_+ \ln (2p_+) + p_- \ln (2p_-).
\end{equation}
Substituting $p_\pm = \tfrac12(1\pm M/M_0)$ and simplifying, we obtain
\begin{equation}
  D_{\rm KL}\!\left(p\Big\Vert\frac{1}{2}\right)
  = \frac{1}{2}
  \bigg[
    \left(1-\frac{M}{M_0}\right)
      \ln \!\left(1-\frac{M}{M_0}\right)
  + \left(1+\frac{M}{M_0}\right)
      \ln \!\left(1+\frac{M}{M_0}\right)
  \bigg].
\label{eq:binary-KL}
\end{equation}
Thus the KL divergence is a simple even function of $M/M_0$ that vanishes at
$M=0$ and increases as the bias away from the uniform distribution grows.

We now consider an ensemble of $N$ such Bernoulli variables, all with the
same bias $p_\pm(M)$.  Assuming independence, the probability distribution
for the full $N$-bit string is the $N$-fold product of the single-bit
distribution, and the KL divergence of this $N$-bit distribution relative to
the corresponding unbiased product distribution is simply
\begin{equation}
  D_{\rm KL}^{(N)}\!\left(p\Big\Vert\frac{1}{2}\right)
  = N\,D_{\rm KL}\!\left(p\Big\Vert\frac{1}{2}\right),
\label{eq:total-KL}
\end{equation}
since relative entropy is additive for independent subsystems.  Using
\eqref{eq:binary-KL}, we can write this more explicitly as
\begin{equation}
  D_{\rm KL}^{(N)}\!\left(p\Big\Vert\frac{1}{2}\right)
  = \frac{N}{2}
  \bigg[
    \left(1-\frac{M}{M_0}\right)
      \ln \!\left(1-\frac{M}{M_0}\right)
  + \left(1+\frac{M}{M_0}\right)
      \ln \!\left(1+\frac{M}{M_0}\right)
  \bigg].
\label{eq:total-KL-explicit}
\end{equation}

Comparing \eqref{eq:total-KL-explicit} with the expression for the
black-hole entropy $S_{\rm bh}(M)$ obtained in Eq.~(\ref{micro})
(or equivalently Eq.~\eqref{eq:Sbh-compact}), we find that 
\begin{equation}
  S_{\rm bh}(M)
  = D_{\rm KL}^{(N)}\!\left(p\Big\Vert\frac{1}{2}\right).
\label{eq:S-as-KL}
\end{equation}
In other words, the black-hole entropy in our model is precisely the
KL divergence between the $N$-bit biased distribution associated with the
black hole mass $M$ and the $N$-bit unbiased reference distribution.

This identification ties together the different viewpoints developed above.
First,  from Eq. \eqref{eq:Sbh-ratio} we see that $e^{S_{\rm bh}}$ is not the number of black-hole microstates, nor does it
count the internal degrees of freedom.  If black hole entropy counts number of internal states has also been asked for the Bekenstein-Hawking entropy \cite{Sorkin:1997ja,Jacobson:1999mi,Sorkin:2005qx,Hsu:2008yi,Engelhardt:2017aux}.
In the present setting, the number of effective microscopic degrees of
freedom is fixed by $N$. The thermodynamic entropy $S_{\rm bh}$ measures,
through its KL representation, how different the black hole distribution is
from the maximally random reference distribution corresponding to $M=0$. In this sense,
$e^{-S_{\rm bh}}$ gives the fraction of the maximally random configurations
that are still allowed after we impose the condition that the system describes a
black hole of mass $M$. Thus $S_{\rm bh}$ measures not an absolute number of
internal states, but the information gained by distinguishing the black-hole
state from empty Minkowski space.

Second, 
this interpretation also gives a natural local geometry to the space of
black-hole ensembles. If the mass is varied infinitesimally,
$M\to M+\delta M$, the corresponding distribution changes from $p_\pm(M)$ to
$p_\pm(M+\delta M)$. For two nearby distributions, the KL divergence has the
expansion
\begin{eqnarray}
D_{\rm KL}^{(N)}\Big(p(M+\delta M)\Big\|p(M)\Big)
=
\frac{1}{2}\,g_{MM}\,\delta M^2
+
O\!\left(\delta M^3\right),
\end{eqnarray}
where 
\begin{equation}
g_{MM}(M)
=
N\sum_{\sigma=\pm}
p_{\sigma}(M)
\left[
\partial_M\log p_{\sigma}(M)
\right]^2
\label{eq:Fisher-definition}
\end{equation}
is the Fisher information associated with the parameter $M$.
Since the Fisher information for $N$ independent  Bernouli trials is $N/p(1-p)$ \cite{Ly:2017Fisher}, we find that 
\begin{eqnarray}
    g_{MM}=\frac{1}{4 M_0^2}\frac{N}{p_+p_-},
\end{eqnarray}
so that 
\begin{eqnarray}
D_{\rm KL}^{(N)}\Big(p(M+\delta M)\Big\|p(M)\Big)
=\frac{1}{8 M_0^2}\frac{N}{p_+ p_-}\delta M^2=
\frac{1}{2 M_0^2}\frac{N}{1-\frac{M^2}{M_0^2}}\delta M^2.
\label{eq:Fisher}
\end{eqnarray}
Thus, the Fisher
information  acts as a local metric and the Fisher–Rao line element~\cite{AmariNagaoka:2000} is 
\begin{eqnarray}
ds_{\rm info}^2=g_{MM} dM^2= \frac{1}{M_0^2}\frac{N}{1-\frac{M^2}{M_0^2}}d M^2
\label{eq:Fisherm}
\end{eqnarray}
on the space of microscopic
probability distributions.
In the infrared limit 
$M_0\gg M$, the local KL divergence between two nearby
black-hole ensembles becomes
\begin{eqnarray}
D_{\rm KL}^{(N)}\Big(p(M+\delta M)\Big\|p(M)\Big)
\approx
\frac{1}{2}\frac{\delta M^2}{M_{\rm Pl}^2},
\label{eq:DKL-approx}
\end{eqnarray}
and thus, the Fisher information obtained from the microscopic KL divergence 
coincides with the Hessian scale of the classical Bekenstein--Hawking entropy \cite{Weinhold:1975xej,Ferrara:1997tw,Ruppeiner:1995zz,Aman:2003ug}.
In this case, the metric \eqref{eq:Fisherm} turns out to be the Weinhold-Ruppeiner metric
\begin{eqnarray}
ds_{\rm info}^2\approx \frac{dM^2}{ M_{\rm Pl}^2}. 
\label{eq:ds-info}
\end{eqnarray}
Therefore, $M_{\rm Pl}$ sets the natural information\! -\! geometric scale
between nearby Schwarzschild macrostates in the infrared limit.

The equality in Eq.~\eqref{eq:Fisherm} is closely related to the
Hessian structures that appear in thermodynamic geometry
\cite{Weinhold:1975xej,Ferrara:1997tw,Ruppeiner:1995zz,Aman:2003ug}.
However, the Fisher--Rao metric obtained here should not be identified
directly with the conventional Weinhold or Ruppeiner metric. In particular,
the Fisher--Rao metric is positive because it arises from a KL divergence,
whereas the usual entropy Hessian of an asymptotically flat Schwarzschild
black hole reflects its negative heat capacity. The important point is that
the positive metric in Eq.~\eqref{eq:Fisherm} follows directly from the
microscopic probability distributions rather than being introduced
phenomenologically.

The Fisher information also determines how accurately the black-hole mass can
be inferred from the microscopic variables. The Cram\'er--Rao inequality~\cite{LehmannCasella:1998} implies
that, for any unbiased estimator $\widehat{M}$ constructed from a single
realization of the $N$-bit configuration,
\begin{equation}
    \operatorname{Var}(\widehat{M})
    \geq
    \frac{1}{g_{MM}(M)}
    =
    \frac{M_0^2}{N}
    \left(
        1-\frac{M^2}{M_0^2}
    \right).
    \label{eq:CR-mass}
\end{equation}
Using $N=M_0^2/M_{\rm Pl}^2$, this becomes
\begin{equation}
    \Delta M
    \equiv
    \sqrt{\operatorname{Var}(\widehat{M})}
    \geq
    M_{\rm Pl}
    \sqrt{
        1-\frac{M^2}{M_0^2}
    }.
    \label{eq:CR-deltaM}
\end{equation}
Therefore, in the infrared regime $M\ll M_0$,
\begin{equation}
    \Delta M \gtrsim M_{\rm Pl}.
    \label{eq:CR-planck}
\end{equation}

Within the present statistical model, the Planck mass therefore sets the
characteristic uncertainty with which the mass $M$ can be inferred from a
single microscopic configuration. This does not mean that two black holes
whose masses differ by $M_{\rm Pl}$ are fundamentally indistinguishable.
Rather, it means that a mass difference of this order is comparable to the
statistical resolution provided by one realization of the microscopic
ensemble.

The same conclusion follows from the local KL divergence. In the infrared
limit, Eq. \eqref{eq:DKL-approx} holds so that
for $\delta M\ll M_{\rm Pl}$, the KL divergence is much smaller than unity,
and the two probability distributions are very close. Consequently, a single
microscopic configuration contains little statistical information with which
to distinguish the two masses. When $\delta M\sim M_{\rm Pl}$, the KL
divergence becomes of order unity, and the two ensembles begin to be
statistically distinguishable. This does not imply perfect discrimination,
but it identifies $M_{\rm Pl}$ as the natural single-sample mass-resolution
scale of the Bernoulli ensemble.

If $K$ independent microscopic configurations are available, the Fisher
information is multiplied by $K$, and the Cram\'er--Rao bound becomes
\begin{equation}
    \Delta M
    \gtrsim
    \frac{M_{\rm Pl}}{\sqrt{K}}
\end{equation}
in the infrared limit. The Planck-scale uncertainty is therefore not an
absolute bound on mass measurements or a statement of mass quantization. It
is the statistical resolution associated with one realization of the
microscopic ensemble assumed in the present model.

\section{Conclusions}

In this work we  revisited the thermodynamics of black holes from the
viewpoint of the third law and information theory.  We started  from the
observation that the standard Bekenstein-Hawking area law does not satisfy
the Nernst formulation of the third law. 
Imposing the latter as an additional thermodynamic requirement, we derived  a
modified entropy
function $S_{\rm bh}(M)$ that approaches the universal residual value $N \ln 2$ at zero temperature, and reduces to the
familiar area law in a suitable limit.   A key ingredient was the  introduction of  
a universal mass scale $M_0$, which  fixes the normalization
of the entropy at low temperatures in much the same way as a minimum vacuum
energy density fixes the absolute normalization of the entropy of black-body
radiation once its high-frequency behaviour is known.  In the limit
$M_0 \to \infty$ the new expression smoothly approaches the Bekenstein-Hawking
entropy, so that the usual area law is recovered as a particular corner of a
broader thermodynamic framework.

A central result of our analysis is the microscopic and
information-theoretic interpretation of the entropy $S_{\rm bh}(M)$.  We have shown that it can be written as the relative entropy, or Kullback–Leibler divergence, between the product distribution of $N$ effective microscopic Bernoulli bits with mass-dependent bias and the uniform distribution over $N$ bits.
  In other words,  the black hole entropy
$S_{\rm bh}$  measures the information-theoretic distance between these
two ensembles.

The representation of the black hole entropy as a KL divergence admits a transparent interpretation in terms of visible
and hidden microstates.  Out of the $W_{\rm tot}=2^N$ possible microstates of $N$ bits,
only a subset of order $W_{\rm vis}=\exp[N H(p)]$ is distinguishable to an exterior
observer who only has access to the biased statistics, where $H(p)$ is the
Shannon entropy of the Bernoulli distribution with bias $p(M)$.  The ratio $W_{\rm tot}/W_{\rm vis}$ characterizes the effective reduction
of the accessible configuration space, and $S_{\rm bh}$ is precisely the
logarithm of this ratio. Since this ratio need not be an integer, it should
be interpreted as an information deficit rather than as an exact
multiplicity of hidden microstates.

Finally, we have shown that the entropy formula admits a natural $1/N$
expansion, with the Bekenstein-Hawking area term as the leading
contribution and subleading terms describing finite-information corrections.
This structure parallels familiar large-$N$ expansions in many-body physics
and suggests a systematic way to organize deviations from the classical area
law in terms of the underlying number of microscopic degrees of freedom.

Several directions for future work follow from our results. On the thermodynamic side, it would be interesting to extend this analysis to charged and rotating black holes, as well as to more general spacetime backgrounds. On the microscopic side, the Bernoulli model used here is intentionally simple. A more complete quantum-gravity description, or a connection with holographic and quantum-information models of black-hole microstates, could help clarify the meaning of the universal mass scale $M_0$ and the importance of finite-$N$ corrections. It would also be useful to study the dynamical consequences of the KL-based entropy, for example in stochastic or random-walk models of black-hole evaporation. Finally, one could ask whether similar modifications, compatible with the third law, appear in other gravitational systems. We hope that the thermodynamic, information-theoretic, and microscopic viewpoints developed here can serve as a useful starting point for future work.


\vskip.5in
\acknowledgments
We would like to thank L. \'Alvarez-Gaum\'e, C. Bachas, G. Dvali, E. Kiritsis, K. Papadodimas and  A. Riotto for discussions and correspondence at various stages of this work.

\vskip.2in
\noindent\\
{\bf\large Appendix}
\vskip.04in

\appendix
\section{Pinsker Inequality and Black Hole Entropy Bound }
\addcontentsline{toc}{section}{Appendix}
The identification of $S_{\rm bh}$ with a KL divergence also allows us to
relate it to a more directly statistical notion of distance from the
reference ensemble.  For any two probability distributions $P$ and $Q$ on
the same discrete space, the (total variation) statistical distance is \cite{Goldreich2001}
defined by
\begin{equation}
  \delta(P,Q)
  = \frac{1}{2}\sum_i \bigl|P_i - Q_i\bigr|.
\end{equation}
In our binary setting, where $P=\{p,1-p\}$ and the reference distribution is
the unbiased Bernoulli law $Q=\{\tfrac12,\tfrac12\}$, this becomes
\begin{equation}
  \delta\!\left(p,\frac12\right)
  = \frac{1}{2}\Bigl(\bigl|p-\tfrac12\bigr|
                     + \bigl|(1-p)-\tfrac12\bigr|\Bigr)
  = \bigl|p-\tfrac12\bigr|.
\label{eq:delta-p-1/2}
\end{equation}
Thus, for a two-outcome distribution the statistical distance to the
uniform distribution is simply the absolute deviation of $p$ from $1/2$.

A general inequality due to Pinsker \cite{Pinsker1964,CoverThomas2006} provides a bound on this statistical
distance in terms of the KL divergence
\begin{equation}
  \delta\!\left(P,Q\right)
  \leq \sqrt{\frac{1}{2}\,D_{\rm KL}(P\Vert Q)}.
\label{eq:Pinsker-general}
\end{equation}
For the Bernoulli case at hand this specializes to
\begin{equation}
  \delta\!\left(p,\frac12\right)
  \leq \sqrt{\frac{1}{2}\,
    D_{\rm KL}\!\left(p\Big\Vert\frac12\right)},
\label{eq:Pinsker-Bernoulli}
\end{equation}
which, upon squaring both sides, can be written as
\begin{equation}
  \delta^2\!\left(p,\frac12\right)
  \leq \frac{1}{2}\,
      D_{\rm KL}\!\left(p\Big\Vert\frac12\right).
\label{eq:Pinsker-squared}
\end{equation}

Using our parametrization of the bias in terms of the black-hole mass,
\begin{equation}
  p = \frac{1}{2}\!\left(1+\frac{M}{M_0}\right),
\end{equation}
Eq.~\eqref{eq:delta-p-1/2} immediately gives
\begin{equation}
  \delta\!\left(p,\frac12\right)
  = \frac{M}{2M_0}.
\label{eq:delta-M/M0}
\end{equation}
Substituting \eqref{eq:delta-M/M0} into the Pinsker bound
\eqref{eq:Pinsker-squared} yields
\begin{equation}
  \left(\frac{M}{2M_0}\right)^{\!2}
  \leq \frac{1}{2}\,
      D_{\rm KL}\!\left(p\Big\Vert\frac12\right),
\end{equation}
or, equivalently,
\begin{equation}
  D_{\rm KL}\!\left(p\Big\Vert\frac12\right)
  \;\geq\;
  \frac{M^2}{2M_0^2}.
\label{eq:KL-lower-bound}
\end{equation}
For the $N$-bit ensemble, the total KL divergence is
\begin{equation}
  D_{\rm KL}^{(N)}\!\left(p\Big\Vert\frac12\right)
  = N\,D_{\rm KL}\!\left(p\Big\Vert\frac12\right)\;\geq\;
  N\,\frac{M^2}{2M_0^2}.
\label{eq:NDKL-lower-bound}
\end{equation}
Since 
$
  N = 8\pi M_0^2, $
the right-hand side of Eq.~\eqref{eq:NDKL-lower-bound} becomes
\begin{equation}
  N\,\frac{M^2}{2M_0^2}
  = 4\pi M^2 =S_{\rm B\text{-}H},
\end{equation}
  and recalling that the microscopic entropy is identified
with the total KL divergence via
\begin{equation}
  S_{\rm bh}(M)
  = D_{\rm KL}^{(N)}\!\left(p\Big\Vert\frac12\right),
\end{equation}
the inequality \eqref{eq:NDKL-lower-bound} can therefore be written as
\begin{equation}
  S_{\rm B\text{-}H}(M)
  \;\leq\;
  S_{\rm bh}(M).
\label{eq:SBH-leq-Sbh}
\end{equation}

Equation~\eqref{eq:SBH-leq-Sbh} has a clear interpretation within our
information-theoretic framework.  Within the Bernoulli realization and the linear bias map adopted here, Pinsker’s inequality implies that the KL entropy is bounded below by its quadratic Bekenstein–Hawking approximation. Our KL-based entropy
$S_{\rm bh}$ quantifies the full information deficit between the biased
ensemble dictated by $M$ and the unbiased reference ensemble. Then, the area law
emerges as a minimal value of this deficit, guaranteed by the universal
Pinsker bound that links relative entropy to statistical distance.  In this
sense, the standard Bekenstein-Hawking result appears as a geometric
lower bound on the information-theoretic distinguishability between the
black-hole ensemble and the maximally mixed state, while our microscopic
construction provides a specific realization that generally saturates
or exceeds this bound.


\bibliographystyle{JHEP}
\bibliography{biblio}

@article{Witten:2024upt,
    author = "Witten, Edward",
    title = "{Introduction to black hole thermodynamics}",
    eprint = "2412.16795",
    archivePrefix = "arXiv",
    primaryClass = "hep-th",
    doi = "10.1140/epjp/s13360-025-06288-y",
    journal = "Eur. Phys. J. Plus",
    volume = "140",
    number = "5",
    pages = "430",
    year = "2025"
}

@article{Bekenstein:1973ur,
    author = "Bekenstein, Jacob D.",
    title = "{Black holes and entropy}",
    doi = "10.1103/PhysRevD.7.2333",
    journal = "Phys. Rev. D",
    volume = "7",
    pages = "2333--2346",
    year = "1973"
}

@article{Bardeen:1973gs,
    author = "Bardeen, James M. and Carter, B. and Hawking, S. W.",
    title = "{The Four laws of black hole mechanics}",
    doi = "10.1007/BF01645742",
    journal = "Commun. Math. Phys.",
    volume = "31",
    pages = "161--170",
    year = "1973"
}

@ARTICLE{BekensteinS,
       author = {{Bekenstein}, Jacob D.},
        title = "{Bekenstein-Hawking entropy}",
      journal = {Scholarpedia},
         year = 2008,
        month = oct,
       volume = {3},
       number = {10},
        pages = {7375},
          doi = {10.4249/scholarpedia.7375},
       adsurl = {https://ui.adsabs.harvard.edu/abs/2008SchpJ...3.7375B}
}

@article{Hawking:1975vcx,
    author = "Hawking, S. W.",
    editor = "Gibbons, G. W. and Hawking, S. W.",
    title = "{Particle Creation by Black Holes}",
    doi = "10.1007/BF02345020",
    journal = "Commun. Math. Phys.",
    volume = "43",
    pages = "199--220",
    year = "1975",
    note = "[Erratum: Commun.Math.Phys. 46, 206 (1976)]"
}

@article{Wald2,
    author = "Wald, Robert M.",
    title = "{The thermodynamics of black holes}",
    eprint = "gr-qc/9912119",
    archivePrefix = "arXiv",
    doi = "10.12942/lrr-2001-6",
    journal = "Living Rev. Rel.",
    volume = "4",
    pages = "6",
    year = "2001"
}

@article{Page:2004xp,
    author = "Page, Don N.",
    title = "{Hawking radiation and black hole thermodynamics}",
    eprint = "hep-th/0409024",
    archivePrefix = "arXiv",
    reportNumber = "ALBERTA-THY-18-04",
    doi = "10.1088/1367-2630/7/1/203",
    journal = "New J. Phys.",
    volume = "7",
    pages = "203",
    year = "2005"
}

@book{wilks1961,
  title     = {The Third Law of Thermodynamics},
  author    = {Wilks, John},
  year      = {1961},
  publisher = {Oxford University Press},
  address   = {Oxford}
}

@article{Wald1,
    author = "Wald, Robert M.",
    title = "{The 'Nernst theorem' and black hole thermodynamics}",
    eprint = "gr-qc/9704008",
    archivePrefix = "arXiv",
    doi = "10.1103/PhysRevD.56.6467",
    journal = "Phys. Rev. D",
    volume = "56",
    pages = "6467--6474",
    year = "1997"
}

@article{KU,
    author = "Kehle, Christoph and Unger, Ryan",
    title = "{Gravitational collapse to extremal black holes and the third law of black hole thermodynamics}",
    eprint = "2211.15742",
    archivePrefix = "arXiv",
    primaryClass = "gr-qc",
    doi = "10.4171/JEMS/1591",
    journal = "J. Eur. Math. Soc.",
    year = "2025",
    note = "Published online first"
}

@article{Reall1,
    author = "Reall, Harvey S.",
    title = "{Third law of black hole mechanics for supersymmetric black holes and a quasilocal mass-charge inequality}",
    eprint = "2410.11956",
    archivePrefix = "arXiv",
    primaryClass = "gr-qc",
    doi = "10.1103/PhysRevD.110.124059",
    journal = "Phys. Rev. D",
    volume = "110",
    number = "12",
    pages = "124059",
    year = "2024"
}

@article{Reall2,
    author = "McSharry, Aidan M. and Reall, Harvey S.",
    title = "{Supersymmetric black holes and the third law of black hole mechanics}",
    doi = "10.1103/s44z-rbzx",
    journal = "Phys. Rev. D",
    volume = "112",
    number = "10",
    pages = "104009",
    year = "2025"
}

@article{Dvali:2017nis,
    author = "Dvali, Gia",
    title = "{Area law microstate entropy from criticality and spherical symmetry}",
    eprint = "1712.02233",
    archivePrefix = "arXiv",
    primaryClass = "hep-th",
    doi = "10.1103/PhysRevD.97.105005",
    journal = "Phys. Rev. D",
    volume = "97",
    number = "10",
    pages = "105005",
    year = "2018"
}

@article{Israel,
    author = "Israel, W.",
    title = "{Third Law of Black-Hole Dynamics: A Formulation and Proof}",
    doi = "10.1103/PhysRevLett.57.397",
    journal = "Phys. Rev. Lett.",
    volume = "57",
    number = "4",
    pages = "397",
    year = "1986"
}

@misc{kehagias-2,
    author = "Kehagias, Alex",
    title = "{Information–Theoretic Black Hole Entropy II: Infrared Gravity and Charged/Rotating Extensions}",
    howpublished = "\textup{({C}ompanion paper)}",
    eprint = "",
}

@article{Dvali2024,
    author = "Dvali, Gia",
    title = "{A string theoretic derivation of gibbons-hawking entropy}",
    eprint = "2407.01510",
    archivePrefix = "arXiv",
    primaryClass = "hep-th",
    doi = "10.1007/s10714-025-03446-6",
    journal = "Gen. Rel. Grav.",
    volume = "57",
    number = "8",
    pages = "118",
    year = "2025"
}

@article{Maldacena:1997de,
    author = "Maldacena, Juan Martin and Strominger, Andrew and Witten, Edward",
    title = "{Black hole entropy in M theory}",
    eprint = "hep-th/9711053",
    archivePrefix = "arXiv",
    doi = "10.1088/1126-6708/1997/12/002",
    journal = "JHEP",
    volume = "12",
    pages = "002",
    year = "1997"
}

@article{Dvali2011,
    author = "Dvali, Gia and Gomez, Cesar",
    title = "{Black Hole's Quantum N-Portrait}",
    eprint = "1112.3359",
    archivePrefix = "arXiv",
    primaryClass = "hep-th",
    doi = "10.1002/prop.201300001",
    journal = "Fortsch. Phys.",
    volume = "61",
    pages = "742--767",
    year = "2013"
}

@article{Strominger1996,
    author = "Strominger, Andrew and Vafa, Cumrun",
    title = "{Microscopic origin of the Bekenstein-Hawking entropy}",
    eprint = "hep-th/9601029",
    archivePrefix = "arXiv",
    reportNumber = "HUTP-96-A002, RU-96-01",
    doi = "10.1016/0370-2693(96)00345-0",
    journal = "Phys. Lett. B",
    volume = "379",
    pages = "99--104",
    year = "1996"
}

@article{Strominger1997,
    author = "Strominger, Andrew",
    title = "{Black hole entropy from near horizon microstates}",
    eprint = "hep-th/9712251",
    archivePrefix = "arXiv",
    reportNumber = "HUTP-97-A106",
    doi = "10.1088/1126-6708/1998/02/009",
    journal = "JHEP",
    volume = "02",
    pages = "009",
    year = "1998"
}

@article{Dvali2015,
    author = {Dvali, Gia and Gomez, Cesar and L{\"u}st, Dieter},
    title = "{Classical Limit of Black Hole Quantum N-Portrait and BMS Symmetry}",
    eprint = "1509.02114",
    archivePrefix = "arXiv",
    primaryClass = "hep-th",
    reportNumber = "MPP-2015-215, LMU-ASC-59-15",
    doi = "10.1016/j.physletb.2015.11.073",
    journal = "Phys. Lett. B",
    volume = "753",
    pages = "173--177",
    year = "2016"
}

@article{Susskind1994,
    author = "Susskind, Leonard and Uglum, John",
    title = "{Black hole entropy in canonical quantum gravity and superstring theory}",
    eprint = "hep-th/9401070",
    archivePrefix = "arXiv",
    reportNumber = "SU-ITP-94-1",
    doi = "10.1103/PhysRevD.50.2700",
    journal = "Phys. Rev. D",
    volume = "50",
    pages = "2700--2711",
    year = "1994"
}

@article{Susskind1994-2,
    author = "Susskind, Leonard",
    title = "{The World as a hologram}",
    eprint = "hep-th/9409089",
    archivePrefix = "arXiv",
    reportNumber = "SU-ITP-94-33",
    doi = "10.1063/1.531249",
    journal = "J. Math. Phys.",
    volume = "36",
    pages = "6377--6396",
    year = "1995"
}

@article{Unruh:1982ic,
    author = "Unruh, W. G. and Wald, Robert M.",
    title = "{Acceleration Radiation and Generalized Second Law of Thermodynamics}",
    doi = "10.1103/PhysRevD.25.942",
    journal = "Phys. Rev. D",
    volume = "25",
    pages = "942--958",
    year = "1982"
}

@article{Zurek:1985gd,
    author = "Zurek, W. H. and Thorne, Kip S.",
    title = "{Statistical mechanical origin of the entropy of a rotating, 
    charged black hole}",
    doi = "10.1103/PhysRevLett.54.2171",
    journal = "Phys. Rev. Lett.",
    volume = "54",
    pages = "2171",
    year = "1985"
}

@article{Frolov:1993fy,
    author = "Frolov, Valeri P. and Page, Don N.",
    title = "{Proof of the generalized second law for quasistationary semiclassical black holes}",
    eprint = "gr-qc/9302017",
    archivePrefix = "arXiv",
    reportNumber = "ALBERTA-THY-10-93",
    doi = "10.1103/PhysRevLett.71.3902",
    journal = "Phys. Rev. Lett.",
    volume = "71",
    pages = "3902--3905",
    year = "1993"
}

@book{GrimmettStirzaker2001ProbabilityRandom,
  author    = {Grimmett, Geoffrey and Stirzaker, David},
  title     = {Probability and Random Processes},
  edition   = {3},
  publisher = {Oxford University Press},
  year      = {2001}
}

@book{Brillouin1956,
  author    = {Brillouin, Leon},
  title     = {Science and Information Theory},
  publisher = {Academic Press},
  address   = {New York},
  year      = {1956}
}

@article{Brillouin1953,
  author  = {Brillouin, Leon},
  title   = {The Negentropy Principle of Information},
  journal = {Journal of Applied Physics},
  volume  = {24},
  number  = {9},
  pages   = {1152--1163},
  year    = {1953},
  doi     = {10.1063/1.1721463}
}

@article{KullbackLeibler1951,
  author  = {Kullback, S. and Leibler, R. A.},
  title   = {On Information and Sufficiency},
  journal = {The Annals of Mathematical Statistics},
  volume  = {22},
  number  = {1},
  pages   = {79--86},
  year    = {1951},
  doi     = {10.1214/aoms/1177729694}
}

@article{Landsberg1984,
  author  = {Landsberg, P. T.},
  title   = {Can Entropy and ``Order'' Increase Together?},
  journal = {Physics Letters A},
  volume  = {102},
  number  = {4},
  pages   = {171--173},
  year    = {1984},
  doi     = {10.1016/0375-9601(84)90934-4}
}

@book{CoverThomas2006,
  author    = {Cover, Thomas M. and Thomas, Joy A.},
  title     = {Elements of Information Theory},
  edition   = {2},
  publisher = {John Wiley \& Sons},
  address   = {Hoboken, NJ},
  year      = {2006}
}

@book{Goldreich2001,
  author    = {Goldreich, Oded},
  title     = {Foundations of Cryptography: Volume 1, Basic Tools},
  edition   = {},
  publisher = {Cambridge University Press},
  address   = {Cambridge},
  year      = {2001},
  isbn      = {0-521-79172-3}
}

@article{Mehra,
	author = {Jagdish Mehra and Helmut Rechenberg},
	doi = {10.1023/a:1018869221019},
	journal = {Foundations of Physics},
	number = {1},
	pages = {91--132},
	publisher = {Kluwer Academic Publishers-Plenum Publishers},
	title = {Planck's Half-Quanta: A History of the Concept of Zero-Point Energy},
	volume = {29},
	year = {1999}
}

@book{Pinsker1964,
  author    = {Pinsker, M. S.},
  title     = {Information and Information Stability of Random Variables and Processes},
  publisher = {Holden-Day Inc.},
  address   = {San Francisco},
  year      = {1964},
  note      = {Translated and edited by Amiel Feinstein}
}

@inproceedings{Sorkin:1997ja,
    author = "Sorkin, Rafael D.",
    title = "{The statistical mechanics of black hole thermodynamics}",
    booktitle = "{Symposium on Black Holes and Relativistic Stars (dedicated to memory of S. Chandrasekhar)}",
    eprint = "gr-qc/9705006",
    archivePrefix = "arXiv",
    month = "5",
    year = "1997"
}

@article{Jacobson:1999mi,
    author = "Jacobson, Ted",
    editor = "Burgess, C. P. and Myers, Robert C.",
    title = "{On the nature of black hole entropy}",
    eprint = "gr-qc/9908031",
    archivePrefix = "arXiv",
    reportNumber = "NSF-ITP-99-094",
    doi = "10.1063/1.1301569",
    journal = "AIP Conf. Proc.",
    volume = "493",
    number = "1",
    pages = "85--97",
    year = "1999"
}

@article{Sorkin:2005qx,
    author = "Sorkin, Rafael D.",
    title = "{Ten theses on black hole entropy}",
    eprint = "hep-th/0504037",
    archivePrefix = "arXiv",
    doi = "10.1016/j.shpsb.2005.02.002",
    journal = "Stud. Hist. Phil. Sci. B",
    volume = "36",
    pages = "291--301",
    year = "2005"
}

@article{Hsu:2008yi,
    author = "Hsu, Stephen D. H. and Reeb, David",
    title = "{Unitarity and the Hilbert space of quantum gravity}",
    eprint = "0803.4212",
    archivePrefix = "arXiv",
    primaryClass = "hep-th",
    doi = "10.1088/0264-9381/25/23/235007",
    journal = "Class. Quant. Grav.",
    volume = "25",
    pages = "235007",
    year = "2008"
}

@article{Engelhardt:2017aux,
    author = "Engelhardt, Netta and Wall, Aron C.",
    title = "{Decoding the Apparent Horizon: Coarse-Grained Holographic Entropy}",
    eprint = "1706.02038",
    archivePrefix = "arXiv",
    primaryClass = "hep-th",
    doi = "10.1103/PhysRevLett.121.211301",
    journal = "Phys. Rev. Lett.",
    volume = "121",
    number = "21",
    pages = "211301",
    year = "2018"
}

@article{Aman:2003ug,
    author = "Aman, Jan E. and Bengtsson, Ingemar and Pidokrajt, Narit",
    title = "{Geometry of black hole thermodynamics}",
    eprint = "gr-qc/0304015",
    archivePrefix = "arXiv",
    reportNumber = "USITP-03-03",
    doi = "10.1023/A:1026058111582",
    journal = "Gen. Rel. Grav.",
    volume = "35",
    pages = "1733",
    year = "2003"
}

@book{Schrodinger1944,
  author    = {Schr{\"o}dinger, Erwin},
  title     = {What is Life? The Physical Aspect of the Living Cell},
  publisher = {Cambridge University Press},
  address   = {Cambridge},
  year      = {1944}
}

@article{Weinhold:1975xej,
    author = "Weinhold, F.",
    title = "{Metric geometry of equilibrium thermodynamics}",
    doi = "10.1063/1.431689",
    journal = "J. Chem. Phys.",
    volume = "63",
    number = "6",
    pages = "2479",
    year = "1975"
}

@article{Ferrara:1997tw,
    author = "Ferrara, Sergio and Gibbons, Gary W. and Kallosh, Renata",
    title = "{Black holes and critical points in moduli space}",
    eprint = "hep-th/9702103",
    archivePrefix = "arXiv",
    reportNumber = "CERN-TH-97-017, CERN-TH-97-17, DAMTP-R-97-09, SU-ITP-97-05",
    doi = "10.1016/S0550-3213(97)00324-6",
    journal = "Nucl. Phys. B",
    volume = "500",
    pages = "75--93",
    year = "1997"
}

@article{Ruppeiner:1995zz,
    author = "Ruppeiner, George",
    title = "{Riemannian geometry in thermodynamic fluctuation theory}",
    doi = "10.1103/RevModPhys.67.605",
    journal = "Rev. Mod. Phys.",
    volume = "67",
    pages = "605--659",
    year = "1995",
    note = "[Erratum: Rev.Mod.Phys. 68, 313--313 (1996)]"
}

@article{Ly:2017Fisher,
  author        = {Ly, Alexander and Marsman, Maarten and Verhagen, Josine
and Grasman, Raoul P. P. P. and Wagenmakers, Eric-Jan},
  title         = {A Tutorial on {Fisher} Information},
  journal       = {Journal of Mathematical Psychology},
  volume        = {80},
  pages         = {40--55},
  year          = {2017},
  doi           = {10.1016/j.jmp.2017.05.006},
  eprint        = {1705.01064},
  archivePrefix = {arXiv},
  primaryClass  = {stat.ME}
}

@book{LehmannCasella:1998,
  author    = {Lehmann, Erich L. and Casella, George},
  title     = {Theory of Point Estimation},
  edition   = {2},
  series    = {Springer Texts in Statistics},
  publisher = {Springer New York},
  address   = {NY},
  year      = {1998},
  doi       = {10.1007/b98854}
}

@book{AmariNagaoka:2000,
  author    = {Amari, Shun-ichi and Nagaoka, Hiroshi},
  title     = {Methods of Information Geometry},
  series    = {Translations of Mathematical Monographs},
  volume    = {191},
  publisher = {American Mathematical Society},
  address   = {Providence, RI},
  year      = {2000}
}

@article{Crump:2026kgu,
    author = "Crump, John R. V. and Gadioux, Maxime and Reall, Harvey S. and Santos, Jorge E.",
    title = "{Violation of the Third Law of Black Hole Mechanics in Vacuum Gravity}",
    eprint = "2601.20955",
    archivePrefix = "arXiv",
    primaryClass = "gr-qc",
    doi = "10.1103/gbg1-pjgq",
    journal = "Phys. Rev. Lett.",
    volume = "136",
    number = "17",
    pages = "171405",
    year = "2026"
}

\end{document}